# Residual strain and piezoelectric effects in passivated GaAs/AlGaAs core-shell nanowires

Moïra Hocevar[1], Le Thuy Thanh Giang[2], Rudeesun Songmuang[2], Martien den Hertog[2], Lucien Besombes[2], Joël Bleuse[3], Yann-Michel Niquet[4], Nikos T. Pelekanos[1,5,6]

[1]CEA, LITEN, INES, 50 avenue du Lac Leman, 73377 Le-Bourget-du-Lac, France

[2]CEA/CNRS/UJF Joint Team "Nanophysics and Semiconductors," Institut Néel-CNRS, BP 166, 25 rue des Martyrs, 38042 Grenoble cedex 9, France

[3]CEA/CNRS/UJF Joint Team "Nanophysics and Semiconductors," CEA/INAC/SP2M, 17 rue des Martyrs, 38054 Grenoble cedex 9, France

[4]Laboratoire de Simulation Atomistique, CEA/INAC/SP2M, 17 rue des Martyrs, 38054 Grenoble cedex 9, France

[5]Department of Materials Science and Technology, University of Crete, P.O. Box 2208, 71003 Heraklion, Greece

[6]Microelectronics Research Group, IESL-FORTH, P.O. Box 1385, 71110 Heraklion, Greece

## Abstract

We observe a systematic red shift of the band-edge of passivated GaAs/Al$_{0.35}$Ga$_{0.65}$As core-shell nanowires with increasing shell thickness up to 100 nm. The shift is detected both in emission and absorption experiments, reaching values up to 14 meV for the thickest shell nanowires. Part of this red shift is accounted for by the small tensile strain imposed to the GaAs core by the AlGaAs shell, in line with theoretical calculations. An additional contribution to this red shift arises from axial piezoelectric fields which develop inside the nanowire core due to Al fluctuations in the shell.

Strain can be used to tune the band structure of semiconductors and thereby engineer their optical and electronic properties. One example is the channel layer of a metal-oxide-semiconductor transistor, where the hole mobility is significantly enhanced by replacing the standard Si channel with strained Si [1]. In compound semiconductors, strain can also induce significant piezoelectric (PZ) fields in polar crystalline orientations [2], especially in the <111> direction of the zinc-blende crystal phase or along the c-axis of hexagonal lattices. Piezoelectric effects are often considered as a burden in III-N semiconductors. For instance, nitride laser diodes typically use large operating currents to counterbalance huge polarization-induced fields [3]. If properly controlled, piezoelectric effects may also benefit semiconductor device properties [4]. In solar cells, they can boost electron-hole pair separation, reducing recombination losses. PZ effects are particularly attractive for III-V nanowire solar cell structures [5,6] because nanowires preferentially grow along the polar <111> direction and are easily strained by growing a shell around them [7-9]. The shell also acts as protective layer for the core surface. For instance, the GaAs surface has an adverse effect on its optoelectronic properties because of the poor quality of its native oxide and the high number of surface traps. This particularly applies to GaAs nanowires, where the high surface-to-volume ratio further increases surface-related effects. *In-situ* epitaxial AlGaAs layers produced prior to GaAs exposure to air provide an efficient passivation scheme. Therefore, GaAs/AlGaAs combinations have long been used in two-dimensional layers, and were lately applied to nanowires [10-14]. In this work, we report on the optical properties of ensembles of GaAs/Al$_{0.35}$Ga$_{0.65}$As core shell nanowires, and we show by a combination of optical spectroscopy and theoretical calculations that the AlGaAs shell not only provides excellent passivation of the GaAs core, but also changes its band structure as a result of the small lattice mismatch between AlGaAs and GaAs.

Self-catalyzed GaAs/AlGaAs core-shell nanowires are grown by molecular beam epitaxy on oxidized Si(111) substrates by the so-called Ga-assisted Vapour-Liquid-Solid (VLS) method, using parameters described in Ref. [15]. We obtain GaAs nanowires with typical heights of 3 μm and an average diameter of 70 nm (Fig. 1a). Their crystal structure is predominantly zinc-blende, with only a few small-size segments consisting of zinc-blende/wurtzite mixtures, as observed under a transmission electron miscoscope (TEM). Prior to shell growth, the Ga droplet is consumed by exposing it to an arsenic flux. The AlGaAs shell is grown at a higher As/Ga ratio in comparison with that used for the GaAs core, to favour two-dimensional growth along the walls of the nanowires (Fig. 1b). We set an average Al content in the shell of 35% by adjusting the relative Al and Ga flux. This average content was independently confirmed by Energy-Dispersive X-ray (EDX) spectroscopy measurements performed along the shells in a scanning electron microscope (SEM). In this study, the nominal shell thickness ranges between 0 and 100 nm (Fig. 1c). On a first sample series the AlGaAs shell is left uncovered, while on a second series, a 5 nm GaAs supershell is grown as a capping layer to ensure that potential oxidation of the AlGaAs shell does not affect our experiments [11]. We do not observe noticeable difference in the results between the two series, with and without the supershell. To investigate optical emission from nanowire ensembles, we perform photoluminescence (PL) measurements using the 405 nm line of a continuous wave (*cw*) laser diode as an excitation source. The PL emission is analyzed with a nitrogen-cooled silicon CCD detector coupled to a monochromator. To study the carrier dynamics, we employ a time-resolved μ-PL setup using a streak camera with an S20 photocathode and a 76-MHz 200-fs pulsed Ti-Sapphire laser. Finally, absorption-type measurements are performed in a micro-photoluminescence excitation (μ-PLE) setup using a *cw* Ti-Sapphire tunable laser to probe the energy band structure of the nanowires.

We determine the optical quality and degree of passivation in our GaAs/AlGaAs core-shell nanowires by studying both the PL intensity and decay time as a function of the shell thickness, keeping the same excitation conditions. Figure 1d shows the spectrally integrated PL intensity at T=5 K from ensembles of nanowires with various shell thicknesses. While bare nanowires have low emission efficiency, we observe a dramatic increase of the intensity with increasing shell thickness. The PL intensity saturates for shells thicker than 20 nm, with an enhancement of over four orders of magnitude compared to bare nanowires. In Figure 1e, the low temperature PL decay time of the band-edge excitonic emission is plotted versus the shell thickness. Here, the decay time increases by more than two orders of magnitude, from the resolution-limited time of 5 ps for core-only nanowires up to 900 ps for 50 nm-thick shells. These results show that our shells play a drastic role in reducing non-radiative recombination channels at the surface of the cores, ensuring lifetimes close to the intrinsic limit [11]. Let us note that the effect of passivation is more efficient in Ga-assisted compared to Au-assisted VLS growth [16], as gold atoms diffuse from the droplet to the lateral surface, acting as non-radiative charge carrier traps and thus degrading the nanowire optical quality.

When looking at the PL spectra in the spectral region of the GaAs core emission, we detect a systematic red shift of the PL peak (Fig. 2a) and a broadening of the PL linewidth with increasing shell thickness, reaching a maximum shift of 14 meV for a shell thickness of 100 nm. The µ-PLE spectra also exhibit a systematic red shift of the absorption band edge (Fig. 2b) with a maximum of 10 meV for a shell thickness of 100 nm, confirming that the shift in PL relates to a real change in the GaAs band structure. The shell thickness dependencies of both the PL emission and absorption band edge (Fig. 2c) strongly suggest a mechanism in which the core is subject to increasing strain as the shell becomes thicker. Furthermore, the increasing Stokes shift between the absorption band edge and the PL peak in Fig. 2c indicates the onset of a weak localization mechanism as the shell thickness increases. Similarly, the PL linewidth increase with shell thickness (Fig.2a) also suggests a shell-dependent disorder mechanism that occurs either within the same nanowire or from one nanowire to the other.

We calculate the strain distribution in our nanowires using a Valence Force Field (VFF) model [17-18] in the Virtual Crystal Approximation for the AlGaAs alloy. The simulated structure consists of a zinc-blende GaAs/Al$_{0.35}$Ga$_{0.65}$As core-shell nanowire with hexagonal cross section, {110} facets and a uniform AlGaAs alloy throughout the shell. The theoretical lattice mismatch between the GaAs core and the AlGaAs shell is $\Delta a/a$ = 0.0486 % for an Al content in the shell of 35%. The length of the nanowire is considered infinite, the core radius $R_c$ is 35 nm, and the shell thickness, $t_s$, varies between 0 and 100 nm. We observe that the largest deformation induced by the shell in the core is $\varepsilon_{zz}$. As discussed in Refs. [8,19], $\varepsilon_{zz}$ is tensile and homogeneous (Fig. 3a). Interestingly, the in-plane deformation $\varepsilon_{||}=(\varepsilon_{xx}+\varepsilon_{yy})/2$ is also tensile, but four times smaller than $\varepsilon_{zz}$ (Fig. 3b). The $\varepsilon_{zz}$ strain behaves as $\varepsilon_{zz} \approx F/(1+F)\, \Delta a/a$, where $F = (t_s^2+2 \cdot R_c \cdot t_s)/R_c^2$, as predicted by a continuum elasticity model [19]. By increasing the shell thickness, $F/(1+F)$ increases rapidly for small $t_s$, and tends to unity for a thick shell (Fig. 3c). For example, $\varepsilon_{zz}$ reaches more than 80% of $\Delta a/a$ for a shell thickness of 50 nm. Next, by using a sp$^3$d$^5$s* tight-binding model [20], we calculate the electronic structure of the system. The single particle bandgap, plotted in Fig. 3d, steadily decreases with increasing shell thickness due to the tensile strain built up in the core. This is in qualitative agreement with the energy shift experimentally observed in both PL and PLE experiments (Fig. 2c).

The model suggests that a large fraction of the observed red shift can be accounted for by mere strain effects. We further investigate if piezoelectric effects also contribute to this shift, by performing excitation power and temperature dependent PL spectroscopy. In Figure 4a, we plot the PL peak energy position as a function of excitation power for various shell thicknesses. Unlike the bare nanowires that exhibit a slight red shift in this power range due to heating effects, we observe a clear blue shift for all core-shell samples. The blue shift increases with excitation power and exhibits a change in slope at around 5 mW. Below 5 mW, the slope is larger for thicker shells and the shift reaches ~5 meV for an excitation power of 5 mW for the 100nm-shell sample. This behavior is typical for piezoelectric nanostructures where carrier-induced screening of the piezoelectric field produces PL blue shifts with increasing excitation power [21-23]. Above 5 mW of excitation power, the screening of the PZ field by photogenerated carriers saturates, while band filling effects take over leading to a stronger blue shift of the PL.

Based on the computed strain fields, we calculate the piezoelectric polarization and potential along the lines of Ref. [18]. The main component of the piezoelectric polarization in the core is $P_z$, whereas the smaller in-plane components give rise only to weak lateral electric fields. The corresponding in-plane piezoelectric potential remains practically constant inside the core even for the thickest shell. Assuming a $\varepsilon_{zz}$ value of 0.05% from Figure 3c, and using the available PZ coefficient for cubic GaAs ($e_{14}$=0.16 C/m$^2$) [24] and the equation A1 from Ref. [25], we find a PZ field oriented mainly along the nanowire axis, with values as high as 8 kV/cm. However, PZ effects are theoretically negligible in a homogeneous and three μm-long nanowire because the associated density of bound charges, $\partial P_z/\partial z$, is non-zero only at the extremities of the nanowire. Consequently, even a small amount of photogenerated carriers is enough to screen the resulting PZ field.

In our case, we experimentally observe PZ related effects with tens of W/cm$^2$ of optical power, suggesting that the PZ fields appear on localization sites with the length scale of the exciton Bohr radius, which makes their photo-screening more difficult and explains the relatively high power densities. This is supported by the fact that the blue shift related to the screening of the PZ field is drastically weakened with increasing temperature and disappears completely at 80 K (Fig. 4b), in line with a weak localization occurring only at low temperatures. This weak localization also explains the temperature dependence of the PL peaks position (Fig. 4c). The PL peak closely follows the empirical Varshni model in bare GaAs nanowires [26]. In contrast, the core-shell nanowires exhibit, at temperatures below 80K, a deviation between the PL peak position and the Varshni curve. Such behavior is characteristic of weak carrier localization occurring in the low temperature range. The deviation increases with shell thickness, reaching ~6 meV for the 100 nm-thick shells, further suggesting that the localization mechanism in the core is directly related to the shell thickness. The presence of a shell-dependent localization mechanism is consistent with the observed increase of the PL linewidth (Fig.2a) and Stokes shift (Fig. 2c) with increasing shell thickness.

Next we discuss the possible origin of the shell-dependent localization mechanism. Towards this end, we apply a combination of TEM imaging and µPL spectroscopy to find out the source of the localization. First, the high-resolution TEM observations on both GaAs and GaAs/AlGaAs nanowires do not bring any evidence of increased structural disorder in the core with increasing shell thickness. We then suggest that the shell-dependent localization mechanism is related to inhomogeneities in the shell along the nanowire axis. Evidence for this is found in µPL measurements (Fig. 5a), where only a few nanowires are probed. There, we see one main PL peak related to band emission from the core. The absence of features in this PL peak suggests that the inhomogeneity occurs within the same nanowire, and not from one nanowire to the other. Moreover, several discrete emission peaks appear above the GaAs energy gap, in the energy range 1.55 to 1.7 eV, *i.e* at energies much lower than the band gap of homogeneous $Al_{0.35}Ga_{0.65}As$ ($E_g$=2 eV). These peaks cannot be attributed to zincblende/wurtzite type II transitions in the shell because their corresponding decay time is too fast (in the order of 200 ps) [27]. We attribute these shell-related emission peaks to inclusions in the shells with lower Al content. To find further evidence for Al inhomogeneity, we then use scanning TEM imaging to study the crystalline structure and the Z-contrast (sensitive to the atomic number of atoms) of the AlGaAs (Fig. 5b-c). Bands of different shades are clearly visible in the top region of the nanowire, showing spontaneously formed Al-rich and Al-poor regions. In this example, we observe ordering on one of the top {110} facets. Similar phase separation is present on all six {110} facets of the nanowires, as confirmed by EDX. Such spontaneous ordering is driven by difference in Ga and Al adatom mobilities on the nanowire surface, leading to Al segregation in the <111> direction [28]. This effect was previously reported in two-dimensional systems as well as in nanowires [28-30]. Such variation in Al content along the shell results in a variation of the strain $\varepsilon_{zz}$ inside the core, and the appearance of PZ fields (Fig. 5d). Consequently, the band structure along the nanowire axis can fluctuate and form weak localization sites, due to the combined action of strain and PZ polarization induced fields. Considering that the core material increasingly feels the strain for the thicker shell, we have a shell-dependent mechanism that can easily generate localization energies of several meV assuming variation of the Al-content along the nanowire axis between 0 and 40%.

In summary, passivated GaAs nanowires with AlGaAs shells exhibit excellent optical properties. The photoluminescence intensity is enhanced by four decades and the carrier decay time is improved by two decades up to 900 ps. We observe a systematic red shift of the band edge of these core-shell nanowires with increasing shell thickness, which can be accounted for by tensile strain imposed on the core by the shell, in line with theoretical calculations of the strain field profile in these composite nanostructures. We argue that piezoelectric fields also contribute to this red shift, acting on localization sites formed inside the nanowire core by inhomogeneities in the Al composition of the shell. On principle, one could take advantage of such PZ fields in future nanowire solar cells to boost electron-hole pair separation.

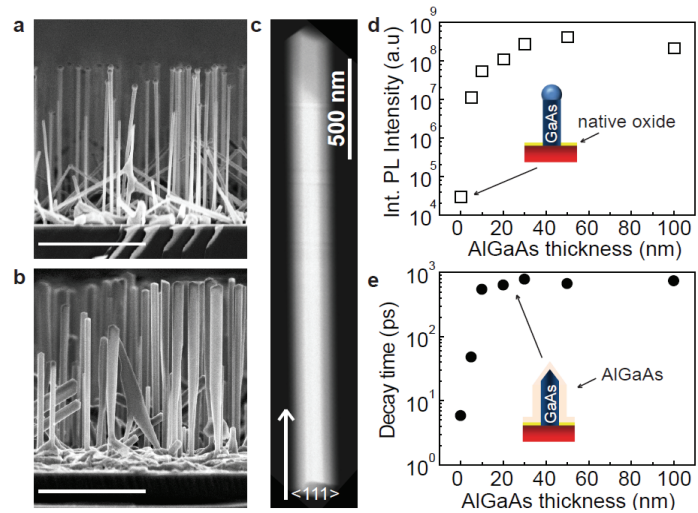

**Figure 1. Passivation of GaAs/AlGaAs nanowires**. SEM cross section images of (a) GaAs and (b) GaAs/AlGaAs core-shell nanowires on Si(111) substrates. Scale bar = 2 μm (c) High Angle Annular Dark Field (HAADF) – scanning TEM (STEM) view of a core-shell nanowire. (d) Integrated photoluminescence intensity versus shell thickness at 5 K. (e) Decay time of the photoexcited carriers versus shell thickness at 5 K.

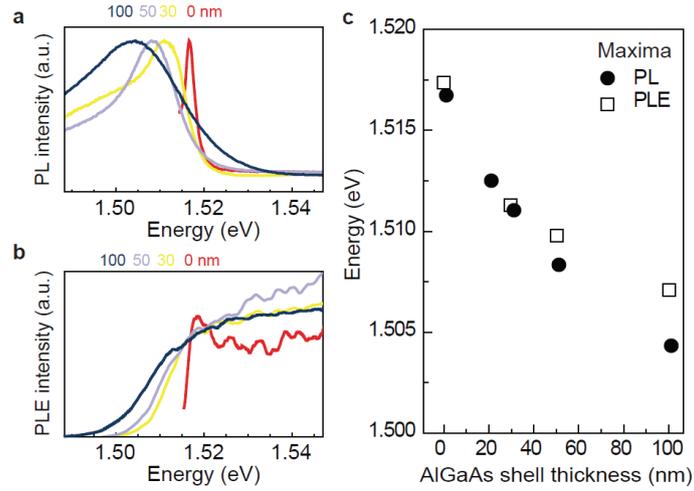

**Figure 2. PL emission and absorption dependence on AlGaAs shell thickness.** (a) 5-K PL spectra of nanowire ensembles with shell thicknesses varying from 0 to 100 nm. (b) 5-K PLE spectra of nanowire ensembles with shell thicknesses varying from 0 to 100 nm. (c) Shift of the band edge emission and absorption photon energy versus shell thickness at 5 K.

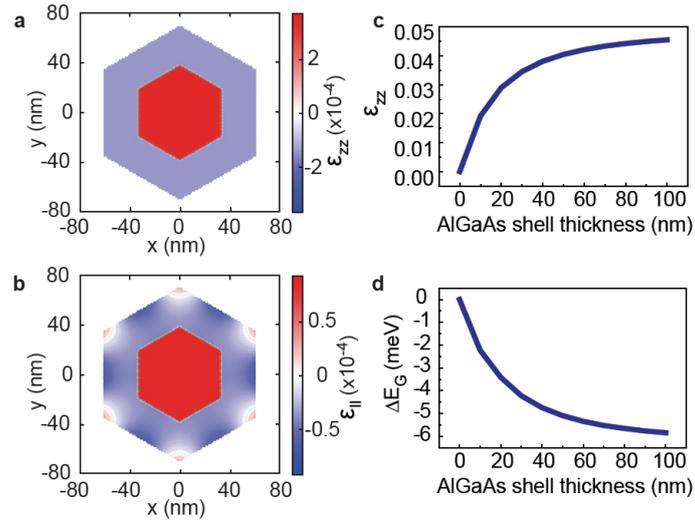

**Figure 3. Theoretical calculations of the residual strain in core-shell GaAs/AlGaAs nanowires**. (a) Cross section view of the axial strain field component $\varepsilon_{zz}$. (b) Cross section view of the average in-plane strain field $\varepsilon_{//}=(\varepsilon_{xx}+\varepsilon_{yy})/2$. The axial strain is much stronger than the in-plane strain. (c) Axial strain in % versus the AlGaAs shell thickness. (d) Calculated strain-induced red shift of the nanowire bandgap versus shell thickness.

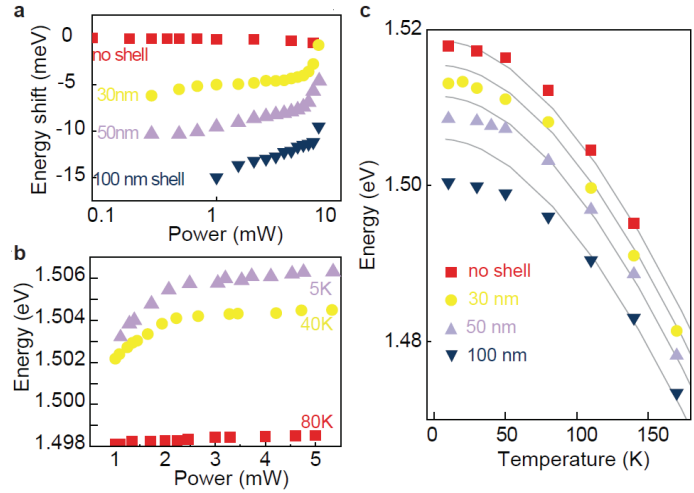

**Figure 4. Experimental evidence of piezoelectricity in GaAs/AlGaAs nanowires.** (a) PL shift versus excitation power at 5K for various shell thicknesses, and (b) PL shift of the 100nm-shell sample for various temperatures. (c) Evolution of the emission band edge versus temperature for various shell thicknesses.

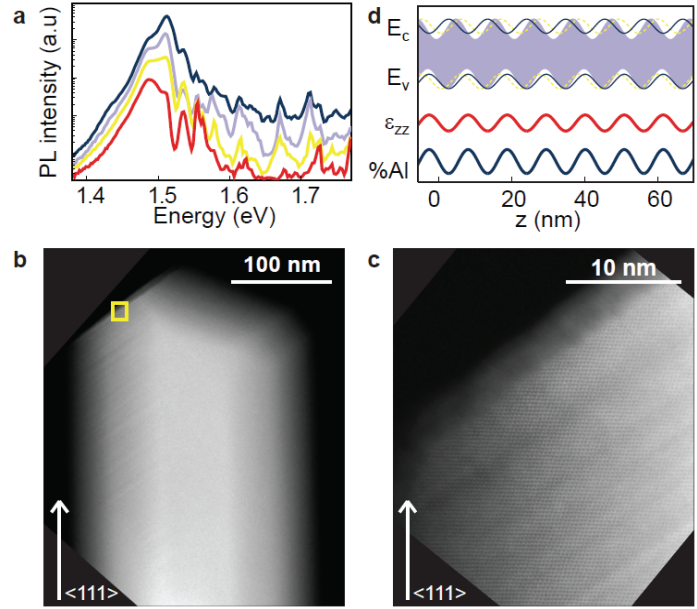

**Figure 5: Inhomogeneity within the AlGaAs shells.** (a) μ-PL at 5 K of few 100 nm-thick core-shell nanowires, where the excitation power increases from P to 4P from bottom to top (P=0.5 mW). The main PL peak at high powers corresponds to the GaAs band edge emission discussed throughout this work. The PL peak at lower energy (1.49eV) is associated to a band-to-acceptor transition involving residual carbon. (b) Nanowire imaged by HAADF-STEM and (c) detail of the shell by high-resolution STEM. (d) Scheme of the total variation of the GaAs band structure (light blue area) along the z-axis due to the simultaneous presence of strain (blue line) and PZ fields (yellow dotted line) induced by variations of the Al content in the shell over length scales comparable to the exciton Bohr radius. Holes and electrons are spatially confined along the nanowire axis.

**Acknowledgements** This work was co–funded by the Institut Carnot Energie du Futur and the European Social Fund and Greek National resources through the THALES program "NANOPHOS". The authors would like to thank Y. Genuist and Y. Cure for the MBE technical support.